\documentclass[%
 aip,
 amsmath,amssymb,
reprint,%
]{revtex4-1}

\usepackage{graphicx}
\usepackage{dcolumn}
\usepackage{bm}
\usepackage{xcolor}
\usepackage[utf8]{inputenc}
\usepackage[T1]{fontenc}
\usepackage{mathptmx}
\usepackage{eurosym}
\usepackage{multirow}
\usepackage{array}
\usepackage{braket}
\usepackage{ragged2e}

\newcolumntype{P}[1]{>{\centering\arraybackslash}p{#1}}

\begin{document}


\title{{\em In itinere} infections covertly undermine localized epidemic control in metapopulations}
\medskip


\author{Francesca Dilisante}%
\affiliation{GOTHAM lab, Institute for Biocomputation and Physics of Complex Systems (BIFI), University of Zaragoza, 50018 Zaragoza (Spain).}
\affiliation{Departament of Condensed Matter Physics, University of Zaragoza, 50009 Zaragoza (Spain).}
\author{Pablo Valga\~n\'on}%
\affiliation{GOTHAM lab, Institute for Biocomputation and Physics of Complex Systems (BIFI), University of Zaragoza, 50018 Zaragoza (Spain).}
\affiliation{Instituto Pirenaico de Ecología – CSIC, Avda. Montañana 1005, 50059 Zaragoza (Spain).}
\author{David Soriano-Pa\~nos}%
\affiliation{Departament de Matem\'aticas i Enginyeria Inform\'atica, Universitat Rovira i Virgili, 43007 Tarragona (Spain).}
\affiliation{GOTHAM lab, Institute for Biocomputation and Physics of Complex Systems (BIFI), University of Zaragoza, 50018 Zaragoza (Spain).}
%
%
\author{Jes\'us G\'omez-Garde\~nes}
\affiliation{GOTHAM lab, Institute for Biocomputation and Physics of Complex Systems (BIFI), University of Zaragoza, 50018 Zaragoza (Spain).}
\affiliation{Departament of Condensed Matter Physics, University of Zaragoza, 50009 Zaragoza (Spain).}
\date{\today}

\begin{abstract}
Metapopulation models have traditionally assessed epidemic dynamics by emphasizing local ({\em in situ}) interactions within defined subpopulations, often neglecting transmission occurring during mobility phases ({\em in itinere}). Here, we extend the Movement–Interaction–Return (MIR) metapopulation framework to explicitly include contagions acquired during transit, considering agents traveling along shared transportation networks. 
We reveal that incorporating {\em in itinere} contagion entails a notable reduction of the epidemic threshold and a pronounced delocalization of the epidemic trajectory, particularly significant in early-stage outbreaks. 
\end{abstract}

\maketitle


\textbf{Recent empirical evidence in the context of the COVID-19 pandemic \cite{Train,subway,TAPIADOR2024105279} indicates that individuals who regularly use mass transportation such as subways, trams, or buses face a significantly higher risk of contracting airborne diseases. In particular, for Influenza-like illnesses \cite{Troko,Gosce}, the occurrence of infections among passengers increases with travel duration and seat proximity, suggesting that higher density and longer journeys amplify the risk of infection. Although many epidemic models incorporate human mobility and travel networks, they typically treat mobility as a single aggregate factor, without explicitly isolating infections acquired during transit.  To address this limitation, we investigate how explicitly modeling the \emph{in itinere} contagion route influences epidemic dynamics compared to frameworks that consider only \emph{in situ} contagions. By leveraging a Markovian metapopulation formalism, we derive a mixing matrix that accounts for both \emph{in situ} and \emph{in itinere} infections in urban contexts. Equipped with this matrix, we show that \emph{in itinere} contagions markedly affect the epidemic threshold and trigger a delocalization transition in early outbreaks—an outcome of paramount importance when designing targeted interventions.}

\section{Introduction}
\label{sec:introduction}

Epidemic modeling has long served as a cornerstone for understanding and mitigating the spread of infectious diseases since the early works by Ross \cite{Ross}, which laid the groundwork for using mathematical approaches such as compartmental models \cite{bailey1957mathematical,anderson1992infectious,wormser2008modeling}. These frameworks, most notably the Susceptible-Infected-Recovered (SIR) model \cite{SIR27}, provided some of the earliest insights into epidemic dynamics, paving the way for practical forecasting tools. However, the inherent mean-field nature of these pioneering approaches overlooked the spatial and behavioral heterogeneities characteristic of real populations, limiting their applicability to qualitative agreement rather than quantitatively precise epidemic trajectories.
\medskip

The advent of network theory \cite{pastor2015epidemic,wang2017unification,annalenMARKOV} and, more specifically, its application to metapopulation dynamics \cite{colizza1,colizza2,PhysRevLett.99.148701,colizza4,tizzoni2014use}, enabled more realistic representations of population structure, wherein individuals interact within localized subpopulations (nodes) and move between them along well-defined mobility patterns. Metapopulation models, especially when coupled with agent-based simulations, have proven instrumental in investigating how human mobility \cite{barbosa2018human} critically shapes epidemic progression, enabling the global spread of disease from localized outbreaks across multiple geographical scales \cite{Eubank,colizza2007epidemic,BALCAN2010132,ajelli2010,tizzoni2012real,SciChinazzi2020}.
\smallskip

When considering metapopulation frameworks amenable to mathematical analysis, different mobility patterns can be modeled by coupling intra-node interactions with inter-node diffusion processes \cite{brockmann2013hidden,Masuda_2010,POLETTO201341,metapopPOL,Castioni2021}. In particular, frameworks incorporating recurrent mobility, mimicking daily commuting, have successfully captured urban epidemics \cite{commutes,prx,gomez2018critical}. Yet, as noted in \cite{BALL201563}, a key challenge lies in accommodating complex social structures beyond the standard assumption that infections arise only \emph{in situ} (within the nodes)—thus neglecting a vital transmission route: \emph{in itinere} contagions, or those acquired during transit. In many real-world scenarios, especially airborne outbreaks such as COVID-19, this additional contagion mechanism \cite{Troko,TAPIADOR2024105279,Train,subway,Gosce} has proven indispensable and, if ignored, may lead to underestimate the epidemic burden, ultimately affecting containment strategies.
\smallskip

In this work, we propose an extension to the metapopulation framework that explicitly incorporates what we refer to as \emph{in itinere contagions}. Building upon the Movement--Interaction--Return (MIR) model \cite{gomez2018critical,Soriano-Panos2018,VBD1,Cota_2021,hazarie2021interplay,soriano2022modeling} with distinguishable agents \cite{DIST,npj-comp}, our approach introduces a third reaction phase accounting for contagions \emph{in itinere}, i.e., while individuals travel between their residential and destination nodes. By extending the mixing matrix formalism \cite{DIST,npj-comp} to cover transit-based contacts, we derive an analytical expression for the epidemic threshold and show that disregarding \emph{in itinere} contagions can overestimate population robustness and mischaracterize the epidemic detriment phenomenon observed in recurrent mobility settings.
\smallskip

\begin{figure*}[t!]
	\centering
		\includegraphics[width=0.95\linewidth]{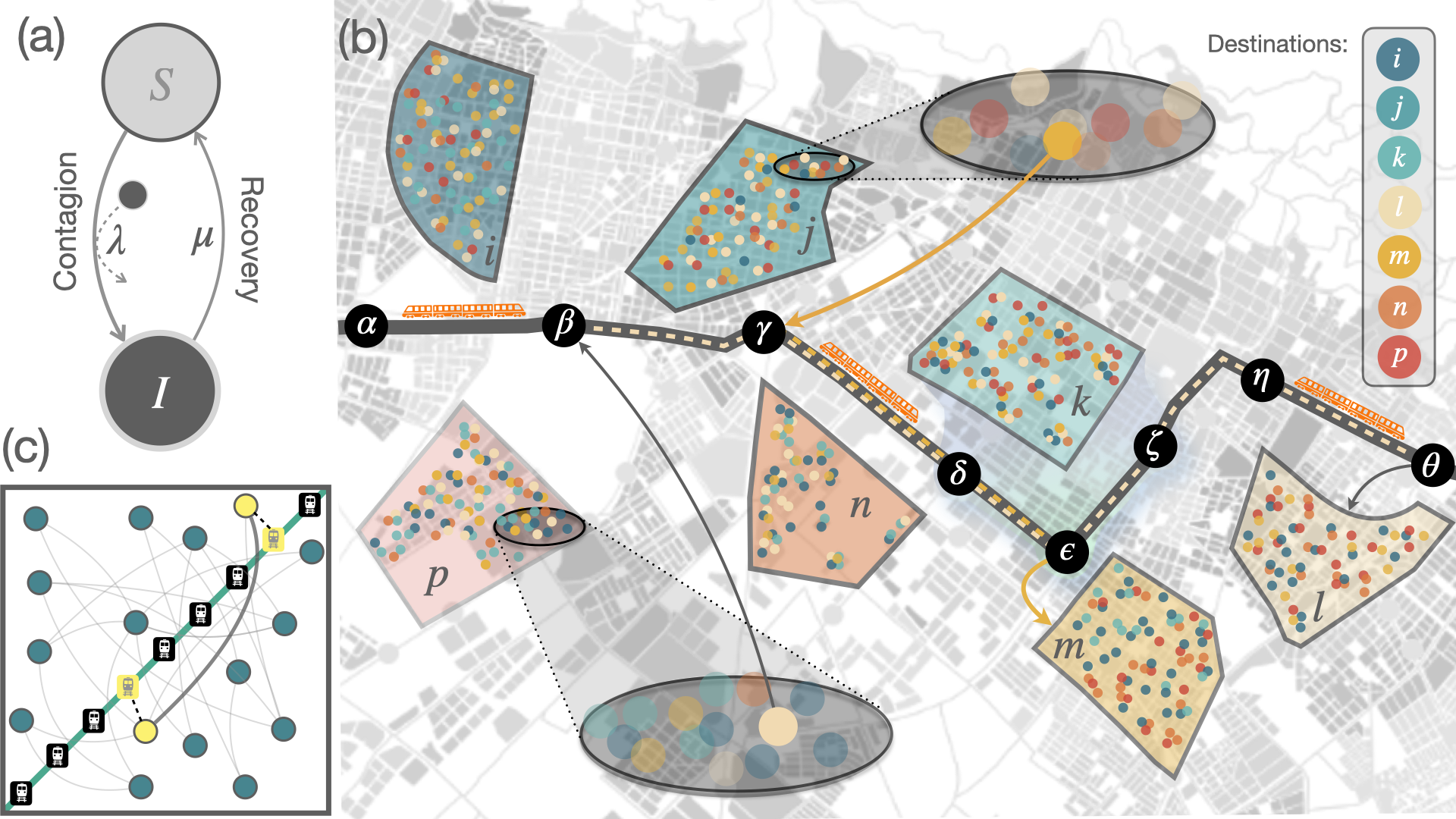}
	\caption{Scheme of the metapopulation framework incorporating {\it in itinere} contagions. (a) Schematic representation of the SIS model. Individuals transition from the susceptible ($S$) to the infected ($I$) state with probability $\lambda$ upon contact with an infectious individual, and recover to the susceptible state with probability $\mu$. (b) Conceptual illustration of the distinguishable-agent MIR framework incorporating \emph{in itinere} contagions. Polygons represent metapopulation nodes (e.g., $i$, $j$, $k$, etc.), with internal dots indicating residents, color-coded by their commuting destination. Black circles ($\alpha$, $\beta$, $\gamma$, …) denote transport stations connected by a transit line (black solid line). Dashed paths illustrate commuting transit routes: one from node $j$ to $m$ (yellow), and another from $p$ to $l$ (cream). Arrows indicate the boarding and alighting stations for each agent, with overlapping segments (e.g., $\gamma \rightarrow \epsilon$) representing shared exposure windows enabling \emph{in itinere} transmission. This mechanism complements standard \emph{in situ} interactions within nodes during day and night phases. (c) Scheme of the synthetic metapopulation model. Nodes are distributed and interconnected within a unit square, while a transport line with equidistant stations spans the diagonal. Boarding and alighting stations are assigned based on proximity between origin and destination nodes and the transport line (e.g. the two stations highlighted in yellow delimit the route followed by individuals commuting between the two nodes highlighted in yellow).}
	\label{fig:1}
\end{figure*}

\section{Model Equations}
\label{sec:model_equations}

In this section, we present the MIR formulation that allows including \emph{in itinere} contagions. To this aim, we first describe the MIR formalism to explain the basic features of the model and then introduce the possibility of contagions during transit. 

\subsection{MIR formalism with distinguishable agents}

In its distinguishable formulation \cite{DIST,npj-comp}, the MIR model categorizes individuals not only by their place of residence but also by their usual destination. To support this formulation one usually relies on the Origin-Destination matrix, ${\bf n}=\{n_{ij}\}$, in which each element $n_{ij}$ captures the total number of commuters from patch $i$ to patch $j$ for the population of interest. This matrix can be viewed as a directed and weighted network with $L$ edges (the nonzero entries of ${\bf n}$). 
\smallskip

Focusing (for simplicity) on the Susceptible-Infectious-Susceptible (SIS) compartmental model (see Fig.\ref{fig:1}.a), the MIR formalism is described by $L \leq N^2$ variables, $\{\rho_{ij}(t)\}$, where $N$ denotes the number of patches in the metapopulation. In particular, each variable $\rho_{ij}(t)$ is the probability that an agent residing in node $i$ and commuting to node $j$ is infectious at time $t$. 
As is common in metapopulation models with recurrent mobility, these $L$ variables $\{\rho_{ij}(t)\}$ evolve in discrete time steps. In particular, the Markovian update equation for an individual living in node $i$ who commutes to node $j$ reads:
\begin{equation}
\rho_{ij}(t+1) \;=\; (1-\mu)\,\rho_{ij}(t)\;+\;\bigl(1-\rho_{ij}(t)\bigr)\,\Pi_{ij}(t),
\label{eq:rho_update}
\end{equation}
where $\mu$ is the SIS recovery probability, and $\Pi_{ij}(t)$ is the probability that a susceptible agent with residence $i$ and destination $j$ becomes infected during the current time step. As detailed below, $\Pi_{ij}(t)$ combines the contagion events that unfold across three sequential processes at each step. In particular, these processes are:
\begin{enumerate}
\item \textbf{Movement (M):}
At the beginning of each time step, individuals are placed in their residence node $i$. Then, with probability $p_d$, an individual travels to node $j$.  Otherwise, with probability $1 - p_d$, the individual remains in node $i$.

\item \textbf{Interaction (I):}
Once at a node (either the residence $i$ or destination $j$), each individual engages in local (\emph{in situ}) contacts that can lead to infection. The contagion probability at node $i$ during this \emph{day} stage is:
\begin{equation}
P^D_i(t) \;=\; 1 \;-\;
\Bigl(
1 \;-\; \lambda\,\frac{I^{\text{eff}}_i(t)}{n^{\text{eff}}_i}
\Bigr)^{z_D\,f_i},
\label{eq:PD}
\end{equation}
where $\lambda$ is the per-contact infection probability of the SIS model, $I^{\text{eff}}_i(t)$ is the effective number of infected agents in node $i$ after movement and
$n^{\text{eff}}_i$ is the node’s effective population. Both quantities are defined as:
\begin{eqnarray}
n^{\text{eff}}_i&=&(1-p_d)\sum_{j}n_{ij}+p_d\sum_{j}n_{ji}\\
I^{\text{eff}}_i(t)&=&(1-p_d)\sum_{j}n_{ij}\rho_{ij}(t)+p_d\sum_{j}n_{ji}\rho_{ji}(t)\;.
\end{eqnarray}
In addition, $z_D = \langle k_D\rangle \sum_i n_i^{eff}/\sum_i f_i n_i^{eff}$ is a scaling factor ensuring an average of $\langle k_D \rangle$ day contacts, and $f_i = n^{\text{eff}}_i / a_i$ encodes the node’s effective density (with $a_i$ denoting its area). Note that agents infected during this day stage remain non-infectious until the following time step.

\item \textbf{Return (R):}
After the interaction phase, all individuals who traveled during the movement stage return to their home node to spend the \emph{night} stage. There, they engage in additional contacts (e.g., household members) that may also result in infections. Being at home, the probability of these {\em in situ} contagions differs from Eq.~(\eqref{eq:PD}):
\begin{equation}
P^N_i(t) \;=\; 1 \;-\;
\Bigl(
1 \;-\; \lambda\,\rho_i(t)
\Bigr)^{\langle k_N \rangle},
\label{eq:PN2}
\end{equation}

where $\rho_i(t)$ is the fraction of infected individuals in node $i$ and $\langle k_N \rangle$ denotes the average number of night contacts in the population (assumed to be equal for all patches). As in the day phase, newly infected agents remain non-infectious until the next time step. Finally, those who were infectious during this time step recover with probability $\mu$, thus entering the next time step as susceptible individuals.
\end{enumerate}

\subsection{Incorporating {\em In Itinere} Contagions}

To capture the risk of infection while traveling, we introduce a third reaction component. When individuals decide to travel, they board a shared transportation network, interacting with other passengers along the way. We assume this network, composed of stations and connecting segments (see Fig.~\ref{fig:1}.b), is unique and universally used by all travellers. Each individual starts at the station nearest their residence and disembarks at the station closest to their destination. The route between these two stations (represented as a sequence of stretches between them) follows the shortest path \cite{dijkstra}. Although certain nodes (such as node $k$ in Fig.~\ref{fig:1}.b) may have multiple equidistant stations, for simplicity, we associate one station with each node.
\smallskip

This expanded distinguishable MIR framework allows for interactions between individuals whose residences and workplaces do not overlap. For instance, consider in Fig.~\ref{fig:1}.b  someone residing in node $p$ and traveling to node $l$, boarding at station $\beta$ and alighting at station $\theta$. This person may interact with another traveler residing in node $j$ and heading to node $m$ while both are simultaneously aboard the transport network, that is, when their routes overlap. Note that, under the baseline MIR model, these individuals would have never encountered one another.
\smallskip

To capture this new transmission route into the mathematical formulation, we write the probability that an individual from node $i$ traveling to node $j$ becomes infected \emph{in itinere} as:
\begin{equation}
P^T_{ij}(t) \;=\; 1 \;-\;
\prod_{(\alpha, \beta)} \Bigl(
1 \;-\; \lambda\,\frac{I(\alpha, \beta)(t)}{n(\alpha, \beta)}
\Bigr)^{c(p_d)\,S_{ij}^{(\alpha, \beta)}},
\label{eq:PT}
\end{equation}
where $S_{ij}^{(\alpha,\beta)}$ is the $(i,j)$ entry of matrix ${\bf S}^{(\alpha,\beta)}$, which encodes whether the path between nodes $i$ and $j$ includes the stretch $(\alpha,\beta)$. Specifically, $S_{ij}^{(\alpha,\beta)}=1$ if the route between $i$ and $j$ uses $(\alpha,\beta)$, and $0$ otherwise. Since the product $\prod_{(\alpha,\beta)}$ runs over all possible station-to-station stretches, the terms $S_{ij}^{(\alpha, \beta)}$ ensure that only those segments actually traveled by $(i \to j)$ affect $P^T_{ij}(t)$. Additionally,
$n(\alpha,\beta)$ denotes the total number of individuals aboard on stretch $(\alpha,\beta)$, which can be explicitly written as:
\begin{equation}
n(\alpha, \beta) \;=\; p_d \, \sum_{k, l} n_{kl} S_{kl}^{(\alpha, \beta)},
\label{eq:nab}
\end{equation}
where $p_d$ scales the sum of all agents $n_{kl}$ whose routes include $(\alpha,\beta)$. Analogously, the number of infected individuals on that same stretch reads:
\begin{equation}
I(\alpha, \beta) (t) \;=\; p_d \, \sum_{k, l} n_{kl} \rho_{kl}(t) S_{kl}^{(\alpha, \beta)}\;.
\label{eq:Iab}
\end{equation}
\smallskip

Each traveler is assumed to make $c(p_d)$ contacts per stretch of the journey. This number is modeled using a first-order Hill equation:
\begin{equation}
c(p_d) \;=\;
c_{sat}\,\frac{p_d}{K + p_d}\;,
\label{eq:chill}
\end{equation}
so that, at low mobility $p_d$, the number of in-transit contacts remains small, whereas at higher mobility it grows asymptotically toward the saturation value $c_{sat}$.
\smallskip

Finally, considering the \emph{in situ} infections during both day (Interaction) and night (Return) phases, along with this \emph{in itinere} infection risk, the overall probability that a susceptible agent resident of node $i$ who commutes to $j$ becomes infected in one time step is:
\begin{align}
\Pi_{ij}(t) &= (1-p_d) \Bigl[ P^D_i(t) + \Bigl(1-P^D_i(t)\Bigr) P^N_i(t) \Bigr]\nonumber \\ &+ p_d \Biggl[ 1 - \Bigl(1-P^T_{ij}(t)\Bigr)\Bigl(1-P^D_j(t)\Bigr) \Bigl(1-P^N_i(t)\Bigr) \Biggr]\;.
\label{eq:Pi_total}
\end{align}
In this expression, the first term corresponds to individuals who do not travel (with probability $1-p_d$) and thus encounter infection only in their residential node $i$. The second term, weighted by $p_d$, applies to those who travel to node $j$, facing potential infections during transit, during the day at the destination, and during the night at home.


\subsection{Mixing matrix and Epidemic threshold}

 The epidemic threshold, $\lambda_c$, is the smallest infection probability for which an endemic regime is attained. Therefore, to derive its value, we focus on the stationary states of the SIS dynamics near $\lambda_c$. In this regime, all variables are time-independent: $\rho_{ij} (t+1) = \rho_{ij}(t) = \rho_{ij}^\star \; \forall \, i, j$. Consequently, Eq.~(\ref{eq:rho_update}) simplifies to:
\begin{equation}
    \mu \rho_{ij}^\star = (1 - \rho_{ij}^\star)\Pi_{ij} (\rho_{ij}^\star).
    \label{eq:rho_uptdate_stat}
\end{equation}
Secondly, for $\lambda \gtrsim \lambda_c$, local prevalences are small: $\rho_{ij}^\star = \epsilon_{ij} \ll 1 \; \forall \, i, j$, enabling the linearization of Eq.~(\ref{eq:rho_uptdate_stat}). The linearized infection probabilities for the day, night, and in-transit reaction processes are:
\begin{eqnarray}
    P_i^D &\approx& p_d \lambda \frac{z^D f_i}{n_i^{eff}} \sum_{j = 1}^N n_{ji}\epsilon_{ji} + \lambda (1-p_d)\frac{z^D f_i}{n_i^{eff}}\sum_{j = 1}^N
 n_{ij} \epsilon_{ij}
 \label{linPD}\\
P_i^N &\approx& \lambda \frac{\langle k_N\rangle}{n_i} \sum_{j = 1}^N n_{ij}\epsilon_{ij}
    \label{linPN}\\
    P_{ij}^T &\approx& \lambda c(p_d)p_d\sum_{k,l} n_{lk}\epsilon_{lk} \sum_{\alpha, \beta}\frac{S_{ij}^{(\alpha
    , \beta)} S_{lk}^{(\alpha, \beta)}}{n(\alpha, \beta)}\;.
    \label{linPT}
\end{eqnarray}
Substituting these expressions into Eq.~(\ref{eq:Pi_total}) yields a linearized form of $\Pi_{ij} (\vec{\rho^\star})$, which, when inserted into Eq.~(\ref{eq:rho_uptdate_stat}) and neglecting nonlinear terms in $\epsilon_{ij}$, leads to the following set of equations for the stationary local prevalence:
\begin{equation}
    \mu \epsilon_{ij}  \approx \Pi_{ij}(\vec{\epsilon}) = \lambda \sum_{k, l}^N M_{jk}^{il}\epsilon_{lk}.
    \label{eq:eigenvalue_problem_1}
\end{equation}
For convenience, in the former expression $\Pi_{ij}(\vec{\epsilon})$ has been expressed as the product of the stationary prevalence vector $\vec{\epsilon}$ and the so-called mixing matrix \cite{DIST,npj-comp}
$\mathbf{M}$, whose element $M_{jk}^{il}$ encodes how individuals with residence $i$ and destination $j$ interact with those with residence $l$ and destination $k$: 
\begin{eqnarray}
M_{jk}^{il} & =  & (1-p_d)p_d\frac{z^D f_k}{n_k^{eff}} n_{lk}\delta_{ik} + (1-p_d)^2 \frac{z^D f_l}{n_l^{eff}} n_{lk}\delta_{il} \nonumber \\ &+& p_d^2 \frac{z^D f_k}{n_k^{eff}} n_{lk}\delta_{jk} + p_d(1-p_d)\frac{z^D f_l}{n_l^{eff}} n_{lk}\delta_{jl} + \frac{\langle k_N\rangle}{n_l} n_{lk}\delta_{il} \nonumber \\ &+&  p_d^2 c(p_d) n_{lk} \sum_{\alpha, \beta} \frac{S_{ij}^{(\alpha, \beta)} S_{lk}^{(\alpha, \beta)}}{n(\alpha, \beta)}.
    \label{eq:mixingmatrix}
\end{eqnarray}
Note that while the first five terms account for {\em in situ} encounters (and correspond to those present in the purely distinguishable-agent mixing matrix framework\cite{DIST}), the last one captures in-transit interactions. 
\smallskip

\begin{figure}[t!]
	\centering
\includegraphics[width=0.75\linewidth]{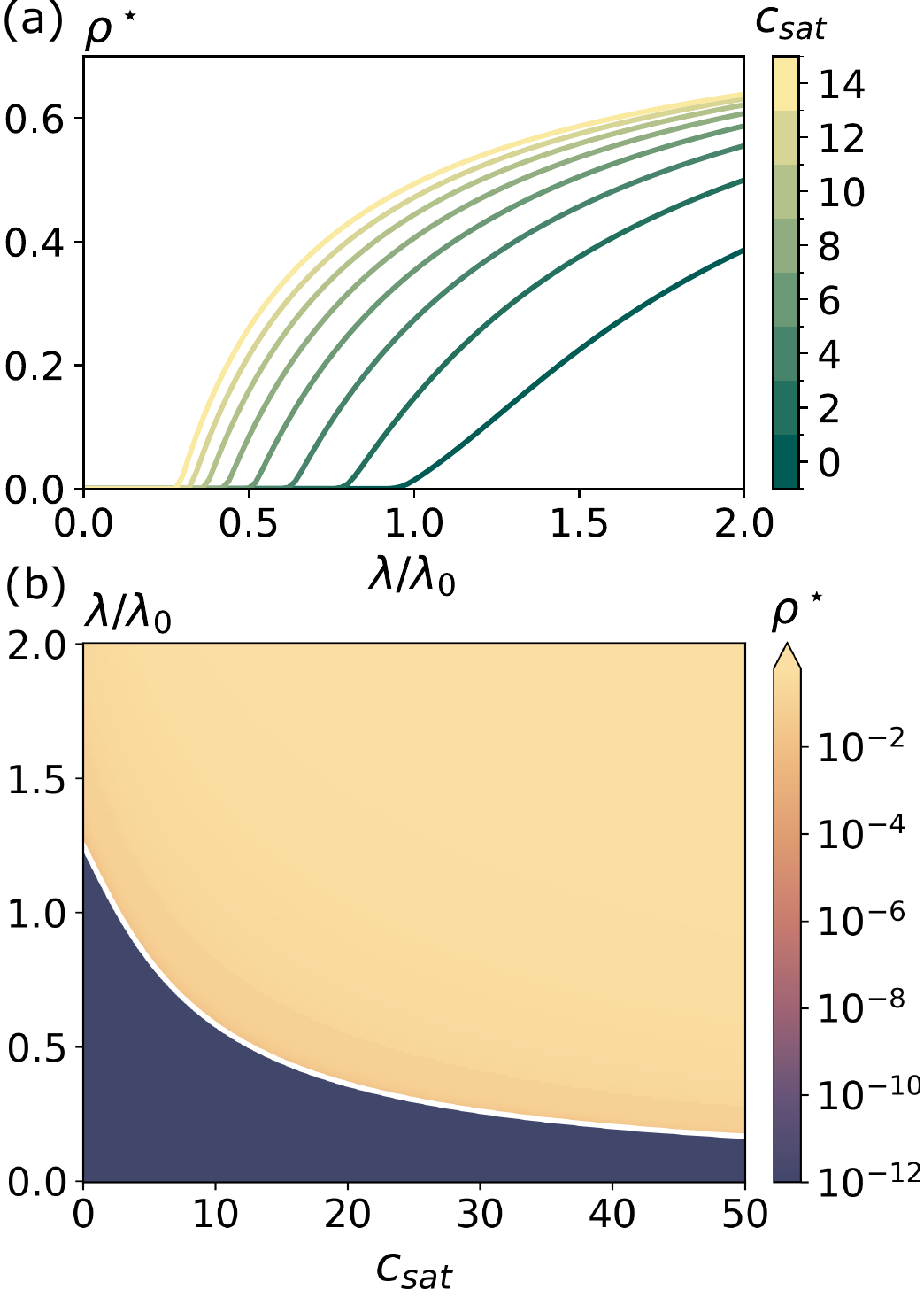}
	\caption{Effect of $c_{sat}$ on epidemic impact and threshold. (a) Steady-state fraction of infected individuals, $\rho^\star$, as a function of the normalized infection probability $\lambda/\lambda_0$, where $\lambda_0$ is the epidemic threshold at $p_d = 0$. Results are shown for $p_d = 0.3$ and various $c_{sat}$ values. (b) Epidemic phase diagram $\rho^\star (\lambda/\lambda_0, c_{sat})$ at $p_d = 0.3$, illustrating the dependence of disease prevalence $\rho^*$ on $\lambda/\lambda_0$ and $c_{sat}$. The solid white line represents the epidemic threshold, $\lambda_c/\lambda_0$, obtained from Eq.~(\ref{eq:threshold}). We have set $K = 10^{-3}$, $\langle k_D\rangle = 8$ and $\langle k_N\rangle = 3$.}
	\label{fig:2}
\end{figure}

Finally, turning our attention to Eq.~(\ref{eq:eigenvalue_problem_1}), it is clear that given ${\bf M}$ this equation can be written as an eigenvalue problem: $(\mu/\lambda)\cdot \vec{\epsilon} = \mathbf{M} \vec{\epsilon}$. Thus, since we are interested in the minimum value of $\lambda$ fulfilling the former expression, the epidemic threshold reads:
\begin{equation}
\lambda_c = \frac{\mu}{\Lambda_{\text{max}}(\mathbf{M})}\;,
\label{eq:threshold}
\end{equation}
where $\Lambda_{\text{max}}(\mathbf{M})$ is the spectral radius of the mixing matrix.

\section{Results}

To assess the impact of {\em in itinere} contagions, we employ a synthetic metapopulation structure with $N = 100$ nodes and average degree $\langle k\rangle= 19$. The nodes have randomly assigned populations averaging $\langle n_i \rangle = 500$, and the edges' weights, $n_{ij}$, are also randomly assigned: a random fraction of each node's population will travel to each of its available destinations. In addition, nodes are spatially distributed within a unit square, interconnected by a transport network composed of $N_T$ = 9 equidistant stations placed along the diagonal, facilitating bidirectional commuting. In Fig.~\ref{fig:1}.c we show a schematic plot of the former metapopulation structure.  
\smallskip

Leveraging this synthetic metapopulation with an integrated transport network, our numerical simulations reveal that explicitly modeling {\em in itinere} infections notably increases the steady-state disease prevalence $\rho^*$, particularly as the in-transit contact parameter  $c_{sat}$ rises (see Fig.~\ref{fig:2}.a). Furthermore, the epidemic threshold $\lambda_c$ significantly decreases with increasing $c_{sat}$, highlighting that ignoring transit-based infections leads to substantial underestimations of the epidemic risk. Analytical predictions derived from Eq.~(\ref{eq:threshold}) align remarkably with numerical results (see Fig.~\ref{fig:2}.b), validating the proposed mixing matrix formulation as a powerful analytical tool for assessing the resilience of populations against infectious disease outbreaks. Note that in this latter plot we use $\lambda_0$, the critical threshold at $p_d = 0$, as a normalization factor for $\lambda$ to focus on the effect of the transit contacts on the threshold. 
\smallskip

\begin{figure}[t!]
	\centering
\includegraphics[width=0.76\linewidth]
{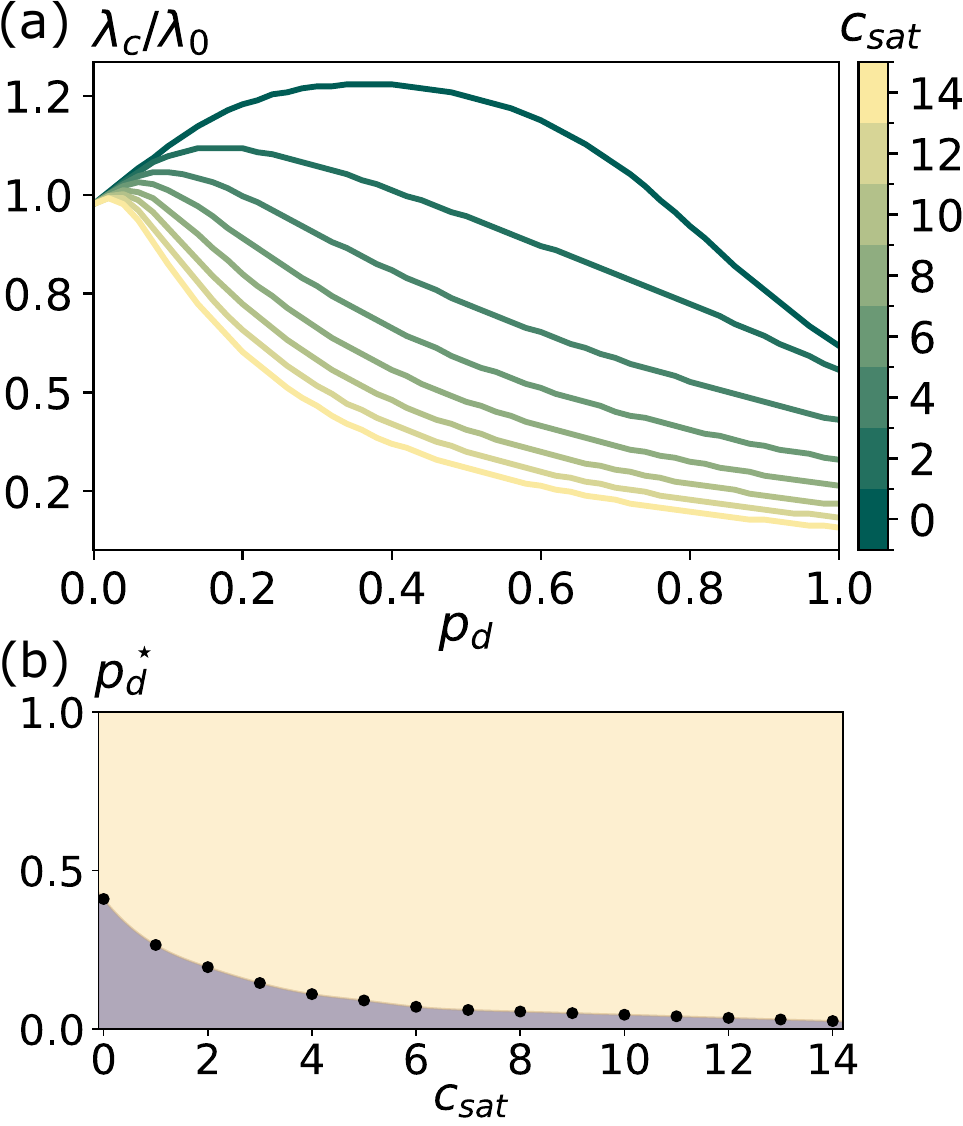}
\caption{Impact of $c_{sat}$ on the epidemic detriment phenomenon. (a) Normalized epidemic threshold $\lambda_c/\lambda_0$ as a function of mobility $p_d$, where $\lambda_0$ denotes the epidemic threshold in the absence of mobility ($p_d = 0$). Results for various $c_{sat}$ values are represented, showcasing how in-transit contacts modulate the critical conditions for epidemic onset. (b) Mobility value $p_d^\star$, defined as the value of $p_d$ that maximizes the epidemic threshold $\lambda_c$, plotted against $c_{sat}$. The shaded regions distinguish regimes where increasing mobility increases $\lambda_c$ (light purple) from those where it decreases it (light yellow). Simulations consider $K = 10^{-3}$, $\langle k_D \rangle = 8$, and $\langle k_N \rangle = 3$.}
\label{fig:3}
\end{figure}


Prior implementations of the MIR formalism have revealed the so-called epidemic detriment phenomenon i.e. an increase in the epidemic threshold, $\lambda_c$, for $p_d \gtrsim 0$ relative to the null mobility scenario ($p_d = 0$).  Although it may initially seem counterintuitive, this effect can be understood by considering the case $p_d = 0$, $\lambda \gtrsim \lambda_0$, where the majority of infections are concentrated within the most vulnerable patch of the metapopulation. When mobility is introduced at a low level, infected individuals can travel to less vulnerable, disease-free patches where the pathogen cannot generate a local outbreak. Concurrently, individuals entering the most affected patch are predominantly healthy, thereby further diluting the disease's prevalence in that patch. As a result, the localized outbreak in the most vulnerable patch can subside without triggering secondary infections in more sparsely populated patches, effectively increasing the epidemic threshold $\lambda_c$. 
\smallskip 

Let us note that, within this formalism, in the null mobility scenario ($p_d = 0$) the most vulnerable patch is the one where most contacts occur per time step: $z^Df_{max} + \langle k_N\rangle$, that is, the one that is most densely populated. In our synthetic metapopulation, all nodes have been assigned the same area, meaning that the most vulnerable patch is the most populated one at low mobility.
\smallskip

Remarkably, the inclusion of {\em in itinere} contagions counteracts the -otherwise seemingly robust- epidemic detriment phenomenon. Figure~\ref{fig:3}.a shows the relationship between $\lambda_c/\lambda_0$ and $p_d$ for increasing values of $c_{sat}$. This figure reveals how in-transit contagion not only reduces the epidemic threshold but also plays a pivotal role in mitigating epidemic detriment. As the number of in-transit contacts increases, this phenomenon is progressively countered, eventually nearly disappearing altogether, as evidenced by the decrease in $p_d^\star$ (the mobility value at which $\lambda_c$ reaches its maximum) with increasing $c_{sat}$ (see Fig.~\ref{fig:3}.b). In the SM, we analytically derive the inverse dependence of $p_d^\star$ through a perturbative analysis of the mixing matrix ${\bf M}$.
\smallskip
\smallskip


To determine the effect of {\em in itinere} contagion on disease localization in the population, we compute the inverse participation ratio (IPR) as a function of mobility, $p_d$, and the per-segment in-transit contacts, $c_{sat}$. The IPR has proven to be a good indicator for localization in spreading dynamics on complex networks\cite{Soriano-Panos2018, IPR_Golstev, IPR_MArtin}. Since we aim to quantify the contribution of each patch to the overall outbreak, we first coarse grain the maximum eigenvector, $\epsilon_{max} (\mathbf{M})$, by summing the contributions associated to each patch $i$, yielding a new eigenvector with $N$ entries, $\mathbf{V}_{max}$:

\begin{equation}
    (\mathbf{V}_{\text{max}})_i = \frac{\sum_{j = 1}^{N} \epsilon^\text{max}_{ij}}{\sqrt{\sum_{k=1}^N \big[\sum_{j = 1}^{N} \epsilon^\text{max}_{ij}\big]^2}}.
    \label{eq:coarse_eigenvect}
\end{equation}

The IPR is then defined as:
\begin{equation}
    \text{IPR} = \sum_{i = 1}^N (\mathbf{V}_{\text{max}})_i^4.
\end{equation}
This quantity is bounded between $1/N$ (fully delocalized, where all patches contribute equally to the epidemic) and $1$ (fully localized, where infections are confined to a single patch).
\smallskip

\begin{figure}[t!]
	\centering
\includegraphics[width=\linewidth]{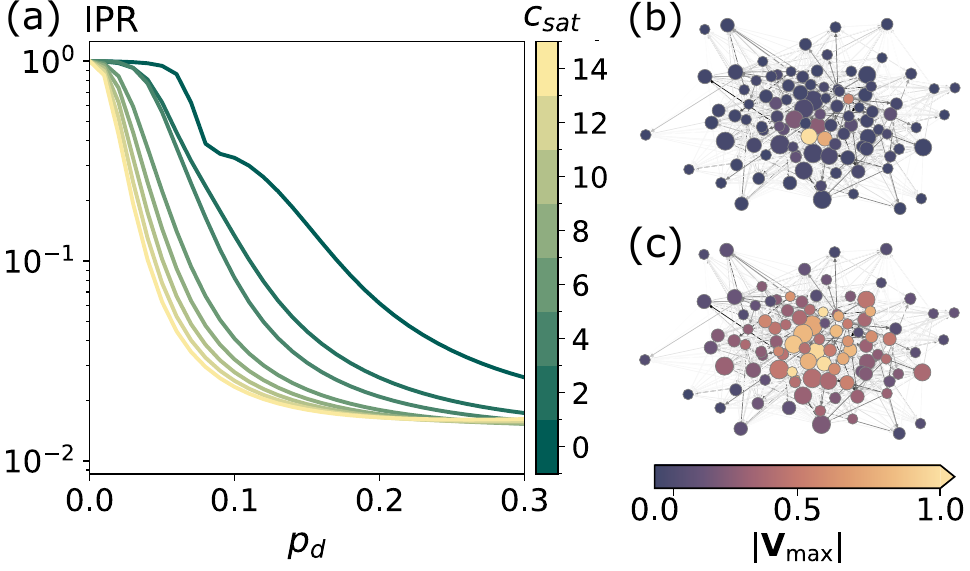}
	\caption{Correlation between $c_{sat}$ and the epidemic delocalization. (a) Inverse Participation Ratio (IPR) as a function of mobility, $p_d$, for several $c_{sat}$ values. (b)-(c) Metapopulation network representations for $p_d = 0.1$ at $c_{sat} = 0$ (panel (b)) and $c_{sat} = 14$ (panel (c)). Node sizes are proportional to the population of each patch, link colors (ranging from white to black) reflect the corresponding $n_{ij}$ values in the OD matrix, and node colors represent the components of the coarse-grained maximum eigenvector $\mathbf{V}_{\text{max}}$, capturing the localization of the infection. Simulations consider $K = 10^{-3}$, $\langle k_D\rangle = 8$ and $\langle k_N\rangle = 3$.}
	\label{fig:4}
\end{figure}

As demonstrated in Fig.~\ref{fig:4}.a, for all values of $c_{sat}$, an increase in $p_d$ results in a decrease in the IPR, pinpointing a delocalization transition driven by geographical mixing due to mobility. Furthermore, as $c_{sat}$ increases, this transition occurs at lower $p_d$, emphasizing the pivotal role of in-transit contacts in driving early delocalization. These contacts contribute to disease homogenization through two main mechanisms: (i) increasing the total number of contacts and (ii) enabling interactions between individuals who neither share their residence nor destination—interactions that do not occur in the baseline MIR model. Additionally, Figs.~\ref{fig:4}.b-c schematically depict the metapopulation, with patches color-coded by their corresponding element in the $\mathbf{V}_{\text{max}}$ vector. This visualization further illustrates that, for fixed mobility ($p_d = 0.1$), increasing in-transit contacts ($c_{sat}$) leads to epidemic delocalization. In this context, $c_{sat}$ governs a transition in the infection distribution from a localized state—confined to a few patches—to a widespread and homogeneous state across the network. Our findings thus represent a cautionary tale for the implementation of targeted strategies, as the delocalization of epidemic states driven by in-transit contagions substantially reduce their suitability as a control strategy to mitigate epidemic outbreaks.

\section{Conclusions}
\label{sec:conc}
In this work, we have proposed a metapopulation framework introducing a novel mechanism that accounts for contagion events occurring during individual transit across a shared transportation infrastructure. By incorporating in itinere contagions into the MIR modeling scheme, we have revealed a non-negligible transmission route that substantially alters both the epidemic threshold and the spatial progression of outbreaks. This refinement provides a more faithful representation of disease dynamics in urban environments, particularly for airborne pathogens.
\smallskip

The analytical derivation of the mixing matrix, which now includes terms representing in-transit contacts, enables a precise characterization of the epidemic threshold through spectral analysis. Numerical simulations, conducted on a synthetic yet structurally realistic metapopulation, confirm that in itinere contagions lead to a significant reduction in the epidemic threshold and an increase in disease prevalence. Moreover, we have shown that the presence of transit-based infections progressively diminishes the epidemic detriment phenomenon, thereby reshaping the critical conditions under which mobility enhances or suppresses epidemic spread.
\smallskip

Beyond the epidemic onset, we have also demonstrated that in itinere contagions are a potent driver of epidemic delocalization. Using the inverse participation ratio as a metric for localization, we have shown that increased transit contacts precipitate a rapid transition from localized to widespread infection states. This finding highlights a fundamental shift in the spatial profile of outbreaks, whereby infections are no longer confined to highly vulnerable patches but become uniformly distributed across the metapopulation.
\smallskip

Overall, our results underscore that ignoring in itinere contagions may lead to significant underestimations of epidemic severity and mischaracterization of spatial dynamics, thereby compromising the design and effectiveness of containment strategies. The framework developed here can be readily adapted to empirical mobility and transportation data, paving the way for more accurate scenario analyses and targeted intervention policies.
\smallskip
\section*{ACKNOWLEDGEMENTS \label{sec:acknowledgements}}
F.D. and J.G.-G. acknowledge support from Departamento de Industria e Innovaci\'on del Gobierno de Arag\'on y Fondo Social Europeo (FENOL group Grant No. E36-23R) and Ministerio
de Ciencia e Innovaci\'on (Grants No. PID2020-113582GB-I00 and No. PID2023-147734NB-I00).  D.S.-P. acknowledges financial support through grants JDC2022-048339-I and PID2021-128005NB- C21 funded by MCIN/AEI/10.13039/501100011033 and the European Union “NextGenerationEU”/PRTR”.


\section*{REFERENCES}

\bibliography{refs}

\renewcommand{\theequation}{S.\arabic{equation}}

\renewcommand{\figurename}{Supplementary Fig.}
\renewcommand{\tablename}{Supplementary Table}

\setcounter{equation}{0}
\setcounter{figure}{0}

\onecolumngrid

\section*{Supplementary material: {\em In itinere} infections covertly undermine localized epidemic control in metapopulations}

\subsection*{A perturbative approach to evaluating the epidemic threshold and analyzing the effect of {\em in itinere} contagion on the epidemic detriment phenomenon.}

The analytical expression for the epidemic threshold, given in Eq. (15) of the main text, requires the knowledge of the largest eigenvalue of the mixing matrix $\mathbf{M}$. Since a closed-form expression for the eigenvalues of $\mathbf{M}$ is not generally available, we compute them via numerical diagonalization. However, in the low-mobility regime ($p_d \approx 0$), perturbation theory provides a valid approximation for the spectral radius of the mixing matrix \cite{gomez2018critical, pert2}, thus enabling the derivation of an analytical expression for the epidemic threshold. This approach offers insight into the behavior of $\lambda_c$ in this regime. Note that, to avoid notational ambiguity, in this supplementary material we denote the epidemic threshold by $\lambda_c$.
\smallskip

Starting from the mixing matrix defined in Eq. (14) of the main text, we reorganize its terms according to powers of $p_d$. Thus, introducing $n'(\alpha, \beta) = \frac{n(\alpha, \beta)}{p_d}$ and $T^{il}_{jk} = \sum_{\alpha, \beta} \frac{S_{ij}^{(\alpha, \beta)}S_{lk}^{(\alpha, \beta)}}{n'(\alpha, \beta)}$ we arrive at:
\begin{equation} 
\begin{split}
(M_{jk}^{il})^T =& \bigg(\frac{z^Df_l}{n_l^{eff}}+ \frac{\langle k_N\rangle}{n_l}\bigg)n_{lk}\delta_{li} + \\
& p_d\Bigg[ \frac{z^Df_k}{n_k^{eff}}\delta_{ki} + \frac{z^Df_l}{n_l^{eff}}\big(-2\delta_{li} + \delta_{lj}\big) + c T^{il}_{jk}\Bigg]n_{lk}+ \\
& p_d^2\Bigg[\frac{z^Df_k}{n_k^{eff}}\big(\delta_{kj}-\delta_{ki}\big) + \frac{z^Df_l}{n_l^{eff}}\big(\delta_{li} + \delta_{lj}\big)\Bigg]n_{lk}.
\end{split}
\label{eq:S1}
\end{equation}

To derive Eq.~(\ref{eq:S1}), we have also considered that the number of in-transit contacts $c (p_d) = c_{sat} \frac{p_d}{K + p_d} \approx c_{sat}$ for $K\ll p_d$, a valid assumption for the values considered in our simulations $K = 10^{-3}$.
\smallskip

Additionally, in order to work with a $N^2 \times N^2$ matrix, we adopt the compact notation $M^{ij}_{lk} = M^{i\cdot N +j}_{l\cdot N+k}$ and, given the decomposition of the mixing matrix into powers of $p_d$, we express it as:
\begin{equation}
\mathbf{M} = \mathbf{M}_0 + p_d \,\mathbf{M}_1 + p_d^2\, \mathbf{M}_2.
\end{equation}

where $\mathbf{M}_0$ represents the unperturbed matrix, and $\mathbf{M}_1$ and $\mathbf{M}_2$ correspond to the first- and second-order perturbative corrections, respectively.
\smallskip





Following perturbation theory, the eigenvalues and eigenvectors of the mixing matrix $\mathbf{M}$ can be expressed as power series in $p_d$:
\begin{eqnarray}
\Lambda^i(p_d) &=& \Lambda_0^i + p_d\,\Lambda_1^i + p_d^2\,\Lambda_2^i + \dots \nonumber \\
\ket{\Psi_i} &=& \ket{0} + p_d\,\ket{1} + p_d^2\,\ket{2} + \dots
\end{eqnarray}
\smallskip

At zeroth order, the eigenvalues are given by:
\begin{equation}
\Lambda_0^i = \Lambda^i(\mathbf{M}_0),
\end{equation}
where $\Lambda^i(\mathbf{M}_0)$ denotes the $i$-th eigenvalue of the unperturbed matrix $\mathbf{M}_0$.
\smallskip

In the present formalism, the structure of $\mathbf{M}_0$ is:
\begin{equation}
(\mathbf{M}_0)^{i\cdot N + j}_{l\cdot N + k} = \left( \frac{z^D f_l}{n_l^{eff}} + \frac{\langle k_N\rangle}{n_l} \right) n_{lk} \delta_{li},
\end{equation}
which is block-diagonal. The blocks are:
\begin{equation}
(\mathbf{M}_0)^i_i = \left( \frac{z^D f_i}{n_i^{eff}} + \frac{\langle k_N\rangle}{n_i} \right)
\begin{pmatrix}
n_{i1} & n_{i2} & \dots & n_{iN} \\
n_{i1} & n_{i2} & \dots & n_{iN} \\
\vdots & \vdots & \ddots & \vdots \\
n_{i1} & n_{i2} & \dots & n_{iN}
\end{pmatrix}.
\end{equation}

This block structure is equivalent to an $N \times N$ matrix with identical rows:
\begin{equation}
\mathbf{A} = 
\begin{pmatrix}
a_1 & a_2 & \dots & a_N \\
a_1 & a_2 & \dots & a_N \\
\vdots & \vdots & \ddots & \vdots \\
a_1 & a_2 & \dots & a_N
\end{pmatrix}.
\end{equation}
\smallskip

Such a matrix has the following well-known spectral properties:
\begin{enumerate}
    \item Its largest eigenvalue is $\Lambda_{\max} = \sum_i a_i$, with associated eigenvector:
    \[
    \ket{v} = \begin{pmatrix} 1 \\ 1 \\ \vdots \\ 1 \end{pmatrix}.
    \]
    \item All other eigenvalues are zero, with a degeneracy of $N - 1$.
\end{enumerate}

\textbf{Proof.} To verify that $\ket{v}$ is an eigenvector of $\mathbf{A}$ with eigenvalue $\Lambda = \sum_i a_i$, we compute:
\[
\mathbf{A} \cdot \ket{v} = 
\begin{pmatrix}
a_1 & a_2 & \dots & a_N \\
a_1 & a_2 & \dots & a_N \\
\vdots & \vdots & \ddots & \vdots \\
a_1 & a_2 & \dots & a_N
\end{pmatrix}
\begin{pmatrix}
1 \\
1 \\
\vdots \\
1
\end{pmatrix}
= \left( \sum_{i=1}^N a_i \right)
\begin{pmatrix}
1 \\
1 \\
\vdots \\
1
\end{pmatrix}.
\]

To confirm that 0 is also an eigenvalue, we consider any vector $\ket{x}$ orthogonal to $\ket{v}$:

\[
\mathbf{A} \cdot \ket{x} = 0,
\]
which implies:
\[
\sum_{j=1}^N a_j x_j = 0.
\]
There exist $N-1$ linearly independent solutions to this equation, forming a basis for the $N-1$-dimensional subspace associated with eigenvalue 0. For instance, we can construct eigenvectors of the form:
\[
\begin{pmatrix}
-1/a_1 \\
1/a_2 \\
0 \\
\vdots \\
0
\end{pmatrix},
\quad
\begin{pmatrix}
-1/a_1 \\
0 \\
1/a_3 \\
\vdots \\
0
\end{pmatrix},
\quad \dots \quad,
\begin{pmatrix}
-1/a_1 \\
0 \\
0 \\
\vdots \\
1/a_N
\end{pmatrix}.
\]
This confirms that the eigenvalue 0 has multiplicity $N-1$, and thus $\Lambda = \sum_i a_i$ is the unique maximum eigenvalue.
\hfill $\blacksquare$
\smallskip

Similarly, the left eigenvectors and eigenvalues of the matrix $\mathbf{A}$ can be obtained as the right eigenvectors and eigenvalues of its transpose, $\mathbf{A}^T$:
\begin{equation}
\mathbf{A}^T = 
\begin{pmatrix}
a_1 & a_1 & \dots & a_1 \\
a_2 & a_2 & \dots & a_2 \\
\vdots & \vdots & \ddots & \vdots \\
a_N & a_N & \dots & a_N
\end{pmatrix}.
\end{equation}
\smallskip

In this case, all columns are identical. Consequently, the matrix $\mathbf{A}^T$ satisfies the following spectral properties:
\begin{enumerate}
    \item Its largest eigenvalue is $\Lambda_{\max} = \sum_i a_i$, with associated eigenvector
    \[
    \ket{v} = \begin{pmatrix}
    a_1 \\
    a_2 \\
    \vdots \\
    a_N
    \end{pmatrix}.
    \]
    \item The only other eigenvalue is 0, with degeneracy $N - 1$.
\end{enumerate}
\smallskip

\textbf{Proof.} We first verify that $\ket{v} = (a_1, a_2, \dots, a_N)^T$ is an eigenvector of $\mathbf{A}^T$ with eigenvalue $\Lambda = \sum_i a_i$:
\[
\mathbf{A}^T \cdot \ket{v} = 
\begin{pmatrix}
a_1 & a_1 & \dots & a_1 \\
a_2 & a_2 & \dots & a_2 \\
\vdots & \vdots & \ddots & \vdots \\
a_N & a_N & \dots & a_N
\end{pmatrix}
\begin{pmatrix}
a_1 \\
a_2 \\
\vdots \\
a_N
\end{pmatrix}
= \left( \sum_{i=1}^N a_i \right)
\begin{pmatrix}
a_1 \\
a_2 \\
\vdots \\
a_N
\end{pmatrix}.
\]

To show that 0 is an eigenvalue, we consider any vector $\ket{x}$ such that:
\[
\mathbf{A}^T \cdot \ket{x} = \vec{a} \cdot \sum_{j=1}^N x_j = \mathbf{0} \quad \Longrightarrow \quad \sum_{j=1}^N x_j = 0.
\]

There exist $N - 1$ linearly independent solutions to this constraint, forming a basis for the subspace corresponding to eigenvalue 0. One possible choice of such eigenvectors is:
\[
\begin{pmatrix}
-1 \\
1 \\
0 \\
\vdots \\
0
\end{pmatrix},
\quad
\begin{pmatrix}
-1 \\
0 \\
1 \\
\vdots \\
0
\end{pmatrix},
\quad \dots \quad,
\begin{pmatrix}
-1 \\
0 \\
0 \\
\vdots \\
1
\end{pmatrix}.
\]

Therefore, the eigenvalue $\Lambda = \sum_i a_i$ is the unique nonzero eigenvalue and corresponds to the principal eigenvector of $\mathbf{A}^T$.
\hfill $\blacksquare$
\smallskip

Applying these results to the mixing matrix blocks, $(\textbf{M}_0)^i_i$, we have that for each block $i$ the largest eigenvalue is:
\begin{equation}
\Lambda_0^i = \bigg(\frac{z^Df_i}{n_i^{eff}}+ \frac{\langle k_N\rangle}{n_i}\bigg) \sum_{j=1}^{N} n_{ij} = \bigg(\frac{z^Df_i}{n_i^{eff}}+ \frac{\langle k_N\rangle}{n_i}\bigg) n_i.
\end{equation}

And therefore the unperturbed matrix, $\mathbf{M}_0$, 's largest eigenvalue will be the largest among them:
\begin{equation}
\Lambda_0^{\max} = \left( \frac{z^D f_i}{n_i^{eff}} + \frac{\langle k_N\rangle}{n_i} \right) n_i \Big|_{\max} = \left( \frac{z^D f_{\max}}{n_{\max}^{eff}} + \frac{\langle k_N\rangle}{n_{\max}} \right) n_{\max}.
\end{equation}

The corresponding right hand side eigenvector's components will all be zero except for those corresponding to the patch with the maximum eigenvalue, which will be 1:
\begin{equation}
 (\ket{0^{\max}})^{i\cdot N + j} = \delta_{i, \max}.
\end{equation}

As for the left-hand side eigenvector, it will read:
\begin{equation}
 \Big(\bra{\Tilde{0}^{\max}}\Big)_{i\cdot N+j} = \frac{1}{n_{\max}}\,\delta_{i, \max}\, n_{ij},
\end{equation}

where $\frac{1}{n_{\max}}$ acts as a normalization factor.
\smallskip

Furthermore, at first and second order, we can calculate the correction of the eigenvalues $\Lambda_1^i$ and $\Lambda_2^i$ as:
\begin{equation}
 \Lambda^i_1 = \bra{\Tilde{0}^i}\mathbf{M}_1\ket{0^i}
\end{equation}

\begin{equation}
\Lambda^i_2 = \bra{\Tilde{0}^{i}}\mathbf{M}_1 \ket{1^{i}} + \bra{\Tilde{0}^i} \mathbf{M}_2 \ket{0^{i}}.
\end{equation}

Where, for a non-degenerated eigenvalue, $\ket{1^i}$ reads:
\begin{equation}
 \ket{1^i} = \sum_{p\neq i} \frac{\bra{\Tilde{0}^p}\mathbf{M}_1\ket{0^i}}{\Lambda_i^0 - \Lambda_p^0} \ket{0^p}.
\end{equation}



Thus, for the largest eigenvalue:
\begin{equation*}
\begin{split}
\bra{\Tilde{0}^{\max}}\mathbf{M}_1\ket{0^{\max}} & = \sum_{i,j} \bra{\Tilde{0}^{\max}}_{i\cdot N+j}(\mathbf{M_1}\ket{0^{\max}})^{i\cdot N+j} \\
 & = \sum_{i, j} \frac{n_{ij}}{n_{\max}}\delta_{i, \max} \bigg( \frac{z^D f_i}{n_i^{eff}} n_{\max\, i} + \frac{z^D f_{\max}}{n_{\max}^{eff}}\big(-2\delta_{\max,i} + \delta_{\max,j}\big)n_{\max} + \\
& \hspace{0.4 cm} + c_{sat} \sum_k n_{\max\,k} T^{i\,\max}_{jk}\bigg) = \sum_j \frac{n_{\max\,j}}{n_{\max}}\bigg( \frac{z^D f_{\max}}{n_{\max}^{eff}} n_{\max\, \max} + \\
 & \hspace{0.4 cm} + \frac{z^D f_{\max}}{n_{\max}^{eff}}\big(-2 + \delta_{\max,j}\big)n_{\max} + c_{sat} \sum_k n_{\max\,k} T^{\max\,\max}_{jk}\bigg) \\
 & = \frac{z^D f_{\max}}{n_{\max}^{eff}} \big( R_{\max, \max} n_{\max} + n_{\max\, \max} - 2 n_{\max}\big) + \\
& \hspace{0.4 cm} + \frac{c_{sat}}{n_{\max}} \sum_{j,k} n_{\max\,j}n_{\max\,k} T^{\max\, \max}_{j\,k} \\
 & = 2 z^D f_{\max} \frac{n_{\max}}{n_{\max}^{eff}} (R_{\max, \max} - 1) + \frac{c_{sat}}{n_{\max}} \sum_{j,k} n_{\max\,j}n_{\max\,k} T^{\max\, \max}_{j\,k}.
\end{split}
\end{equation*}

Where $R_{\max, \max} = n_{\max\,\max}/n_{\max}$ is the \textit{autoloop} fraction at the most vulnerable patch. Note that the first term corresponds to the first-order correction to the largest eigenvalue when in-transit contagion is not considered.
\smallskip

As for the second-order correction to the largest eigenvalue, we have that:
\begin{equation}
\bra{\Tilde{0}^p}\mathbf{M}_1\ket{0^{\max}} = \frac{z^D f_p}{n_p^{eff}} n_{\max \, p} + \frac{z^D f_{\max}}{n_{\max}^{eff} } (-2\delta_{\max \, p}) + \frac{z^D f_{\max}}{n_{\max}^{eff}}\frac{n_{p\, \max}}{n_p} + \frac{c_{sat}}{n_p}\sum_{j, k} n_{pj} n_{\max\, k} T^{p\, \max}_{jk}.
\end{equation}

\begin{equation}
\bra{\Tilde{0}^{\max}}\mathbf{M}_1\ket{0^{p}} = \frac{z^D f_{\max}}{n_{\max}^{eff}} n_{p \, \max} + \frac{z^D f_{p}}{n_{p}^{eff} } (-2\delta_{\max \, p}) + \frac{z^D f_{p}}{n_{p}^{eff}}\frac{n_{\max\, p}}{n_{\max}} + \frac{c_{sat}}{n_{\max}}\sum_{j, k} n_{\max\, j} n_{p k} T^{\max\, p}_{jk}.
\end{equation}

\begin{equation}
\bra{\Tilde{0}^{\max}}\mathbf{M}_2\ket{0^{\max}} = \frac{z^D f_{\max}}{n_{\max}^{eff}} n_{\max} + \sum_j \frac{z^D f_j}{n_j^{eff}} \frac{(n_{\max\, j})^2}{n_{\max}}.
\end{equation}

All expressions derived above can be further simplified by considering that, for the values of $p_d$ at which the perturbative expansion (to second order) remains valid, we can approximate $n_i \approx n_i^{eff}$. Under this assumption, the terms in the expansion of the spectral radius of the mixing matrix become:

\begin{equation}
\Lambda_0^{\max} \approx z^D f_{\max} + \langle k_N\rangle,
\label{eq:correction0}
\end{equation}

\begin{equation}
\Lambda_1^{\max} \approx 2 z^D f_{\max} (R_{\max, \max} - 1) + \frac{c_{sat}}{n_{\max}} \sum_{j,k} n_{\max\,j} n_{\max\,k} T^{\max\,\max}_{j\,k},
\label{eq:correction1}
\end{equation}

\begin{equation}
\begin{split}
\Lambda_2^{\max} \approx & \sum_{p \neq \max} 
\frac{
\left[
\frac{z^D f_p}{n_p} n_{\max\,p} + \frac{z^D f_{\max}}{n_{\max}} \frac{n_{p\,\max}}{n_p} + \frac{c_{sat}}{n_p} \sum_{j,k} n_{p\,j} n_{\max\,k} T^{p\,\max}_{j\,k}
\right]
}{
z^D(f_{\max} - f_p) + z^N (\sigma_{\max} - \sigma_p)
} \\
&\times
\left[
\frac{z^D f_{\max}}{n_{\max}} n_{p\,\max} + \frac{z^D f_p}{n_p} \frac{n_{\max\,p}}{n_{\max}} + \frac{c_{sat}}{n_{\max}} \sum_{j,k} n_{\max\,j} n_{p\,k} T^{\max\,p}_{j\,k}
\right] \\
&+ z^D f_{\max} + \sum_{j} \frac{z^D f_j}{n_j} \frac{(n_{\max\,j})^2}{n_{\max}}.
\end{split}
\label{eq:correction2}
\end{equation}

\begin{figure*}[t]
\centering\includegraphics[width=.55\linewidth]{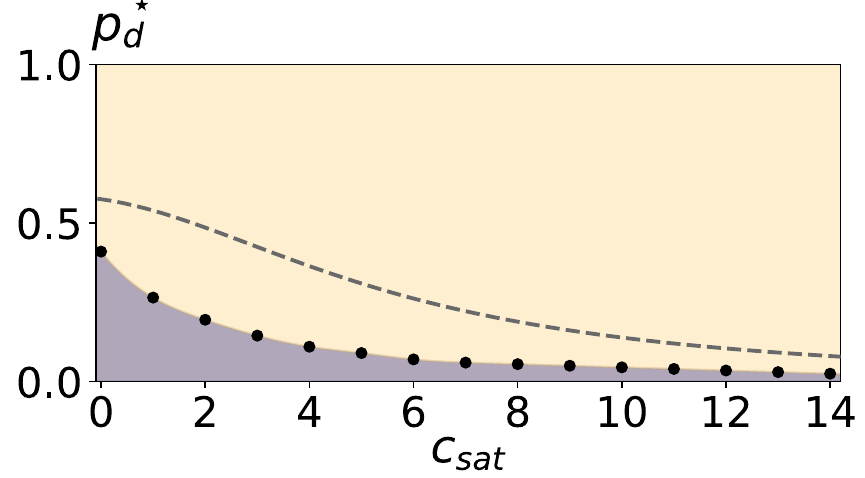}
\caption{\justifying Impact of $c_{sat}$ on epidemic detriment: comparison between numerical and perturbative results. The mobility value $p_d^\star$, defined as the value of $p_d$ that maximizes the epidemic threshold $\lambda_c$, is plotted as a function of $c_{sat}$. Numerical results are shown as black dots, while the dashed line corresponds to the prediction from the perturbative expression given in Eq.~(\ref{eq:pdstar_pert}). Shaded regions indicate distinct dynamical regimes: in light purple, increasing mobility enhances the epidemic threshold, whereas in light yellow, mobility reduces $\lambda_c$. Simulations are performed with parameters $K = 10^{-3}$, $\langle k_D \rangle = 8$, and $\langle k_N \rangle = 3$.}
\label{fig:SM_1}
\end{figure*}

This perturbative analysis of the epidemic detriment phenomenon in the low-mobility regime, reveals that to first order, the detriment is primarily driven by the reduction in contacts between residents of the most vulnerable patch. Furthermore, the positive term in Eq.~(\ref{eq:correction0}) shows that in-transit contagion counteracts epidemic detriment by increasing interactions between residents of the most vulnerable node. Importantly, this counter-detriment effect emerges at first order when assuming $c(p_d) \approx c_{sat}$ as we have done here, and at second order when a Hill function is used to model {\em in itinere} contacts.
\smallskip

Having now an explicit expression for the spectral radius of the mixing matrix, we can derive a fully analytical expression for the epidemic threshold. Furthermore, by differentiating this expression and setting the derivative to zero, we obtain an analytical estimate for $p_d^\star$, the mobility level at which the epidemic threshold reaches its maximum. This marks the turning point beyond which the phenomenon of epidemic detriment begins to subside:

\begin{equation}
\frac{\partial \beta_c}{\partial p_d} = \frac{\partial }{\partial p_d} \left( \frac{\mu}{\Lambda_0^{\max} + p_d \Lambda_1^{\max} + p_d^2 \Lambda_2^{\max}} \right) = 0 
\quad \Longrightarrow \quad 
p_d^\star = \frac{-\Lambda_1^{\max}}{2\Lambda_2^{\max}}.
\label{eq:pdstar_pert}
\end{equation}

To round off the analysis, Supplementary Fig.~\ref{fig:SM_1} presents a comparison between the values of $p_d^\star$ obtained numerically and those predicted analytically by the perturbative expression in Eq.~(\ref{eq:pdstar_pert}). As expected, the agreement improves for higher values of $c_{\text{sat}}$, which corresponds to a weaker epidemic detriment effect and lower values of $p_d^\star$. In such regimes, the system remains closer to the assumptions underlying the perturbative approach—particularly the dominance of the zeroth- and first-order terms—thus enhancing the accuracy of the analytical prediction.

\end{document}